# MANIFOLD DAMPING OF WAKEFIELDS IN HIGH PHASE ADVANCE LINACS FOR THE NLC

R.M. Jones[†], Z. Li[†], R.H. Miller[†], T.O. Raubenheimer[†], R.D. Ruth[†],
G.V. Stupakov[†], J.W. Wang[†], N.M. Kroll[§]

[†]Stanford Linear Accelerator Center,
2575 Sand Hill Road, Menlo Park, CA, 94025
[§]University of California, San Diego
La Jolla, CA 92093-0319

Abstract

Earlier RDDS (Rounded Damped Detuned Structures) [1,2], designed, fabricated and tested at SLAC, in collaboration with KEK, have been shown to damp wakefields successfully. However, electrical breakdown has been found to occur in these structures and this makes them inoperable at the desired gradient. Recent results [3] indicate that lowering the group velocity of the accelerating mode reduces electrical breakdown events. In order to preserve the filling time of each structure a high synchronous phase advance (150 degrees as opposed to 120 used in previous NLC designs) has been chosen. Here, damping of the wakefield is analyzed. Manifold damping and interleaving of structure cell frequencies is discussed. These wakefields impose alignment tolerances on the cells and on the structure as a whole. Tolerance calculations are performed and these are compared with analytic estimations.

*Paper presented at the 2002 8$^{th}$ European Particle Accelerator Conference (EPAC 2002)
Paris, France,
June 3$^{rd}$ -June 7$^{th}$, 2002*

This work is supported by Department of Energy grant number DE-AC03-76SF00515† and DE-FG03-93ER40759§

# MANIFOLD DAMPING OF WAKEFIELDS IN HIGH PHASE ADVANCE LINACS FOR THE NLC

R.M. Jones[†], Z. Li[†], R.H. Miller[†], T.O. Raubenheimer[†], R.D. Ruth[†],
G.V. Stupakov[†], J.W. Wang[†]; SLAC, N.M. Kroll[§]; UCSD

## Abstract

Earlier RDDS (Rounded Damped Detuned Structures) [1,2], designed, fabricated and tested at SLAC, in collaboration with KEK, have been shown to damp wakefields successfully. However, electrical breakdown has been found to occur in these structures and this makes them inoperable at the desired gradient. Recent results [3] indicate that lowering the group velocity of the accelerating mode reduces electrical breakdown events. In order to preserve the filling time of each structure a high synchronous phase advance (150 degrees as opposed to 120 used in previous NLC designs) has been chosen. Here, damping of the wakefield is analyzed. Manifold damping and interleaving of structure cell frequencies is discussed. These wakefields impose alignment tolerances on the cells and on the structure as a whole. Tolerance calculations are performed and these are compared with analytic estimations.

## 1. INTRODUCTION

The previous series of R/DDS have been shown to successfully damp the long-range transverse dipole wakefield in 1.8 meter long structures, each of which consist of 206 cells [2]. The wakefield was measured and found to be well predicted by the spectral function method. However, these structures suffer from electrical breakdown and for this reason the accelerating structures have been redesigned. Particular attention has been paid to reducing the surface field on the cavities, improving surface cleanliness, reducing the overall structure length, and on reducing the group velocity of the fundamental mode. This has resulted in a significant reduction of the number of breakdown events recorded [3]. In this paper we report on manifold damping and detuning of the wakefield in H60VG3. This 55 cell structure is 60 cm in length and has an average group velocity of 0.03c. In performing this design we note that we are restricted to limited couplings to the manifold as the slots sizes are required to be as small as is practical in order to limit the effects of pulse heating in this area. Also, the structures have been optimised to achieve maximum efficiency and thus any change in the central frequency is required to be small in order to preserve the efficiency as much as is possible. The damping of these new structures are reported on section 2 and the alignment tolerances that are imposed on the structures by this wakefield for a specified emittance dilution are discussed in section 3.

## 2. INTERLEAVING OF STRUCTURES

The spectral function code that we use to calculate the wake for this design has recently been highly optimised using sparse matrix techniques. The matrix describing two-fold interleaving of structures is obtained from the non-interleaved single structure circuit model [1,4] which is reformulated in the form:

$$M \begin{pmatrix} a \\ \hat{a} \\ A \end{pmatrix} = \begin{pmatrix} 0 \\ B \\ 0 \end{pmatrix} \quad (2.1)$$

where the elements $a$, $\hat{a}$ and $A$ are themselves column vectors of dimension $2xN$ (= number of cell in a given structure) and M, is the matrix which describes the coupling of modes within the interleaved structures:

$$\begin{pmatrix} f^2 H_1 - 1 & 0 & f^2 H_{1x} & 0 & -f^2 G_1 & 0 \\ 0 & f^2 H_2 - 1 & 0 & f^2 H_{2x} & 0 & -f^2 G_2 \\ f^2 H_{1x}^t & 0 & f^2 \hat{H}_1 - 1 & 0 & 0 & 0 \\ 0 & f^2 H_{2x}^t & 0 & f^2 \hat{H}_2 - 1 & 0 & 0 \\ -f^2 G_1 & 0 & 0 & 0 & f^2 R_1 & 0 \\ 0 & -f^2 G_2 & 0 & 0 & 0 & f^2 R_2 \end{pmatrix}$$

and B describes coupling to the particle beam. Here, the additional subscript refers to the particular structure, interleaved with its neighbour.

In all of our previous calculations [5] of interleaving of the cells of the frequencies of adjacent accelerating structures we have taken the average of each spectral function and from this obtained the overall wakefield. Here, we make a calculation on the global matrix, which for 3-fold interleaving of structures is of a similar form as eq. 2.1 (but of overall dimensions 9xN). We then compare the averaging method with the full matrix method described by 2.1 and we find the difference in the two methods is at worst no more than 2%. This gives us confidence in applying the averaging technique in subsequent calculations; which is significantly more efficient to apply in practice. However, the global matrix technique will be used in the final section in which an analytic method is used to approximate structure tolerances.

The spectral function and wakefield for a manifold damped version of H60VG3 are shown in Figs. 1 and 2 for a single structure and for a three-fold interleaved structure. In all previous structures the wakefield was

[†]Supported by DOE grant number DE-AC03-76SF00515
[§]Supported by DOE grant number DE-FG03-93ER40759

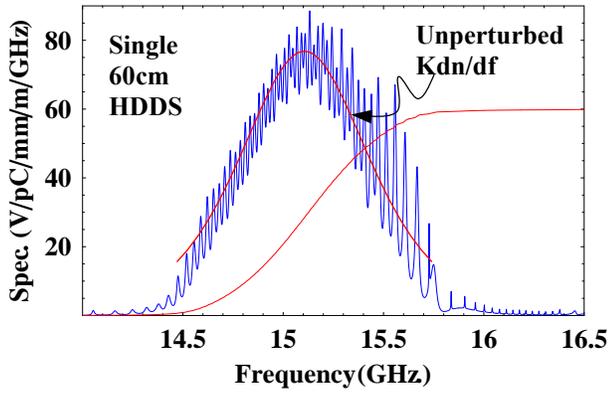
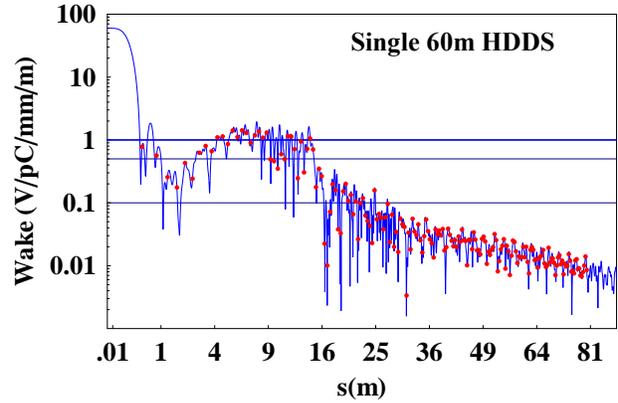
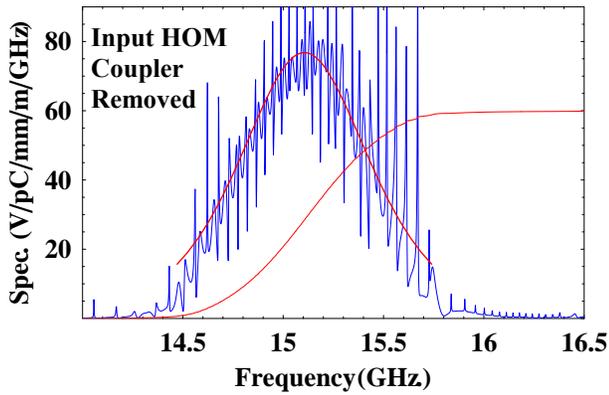
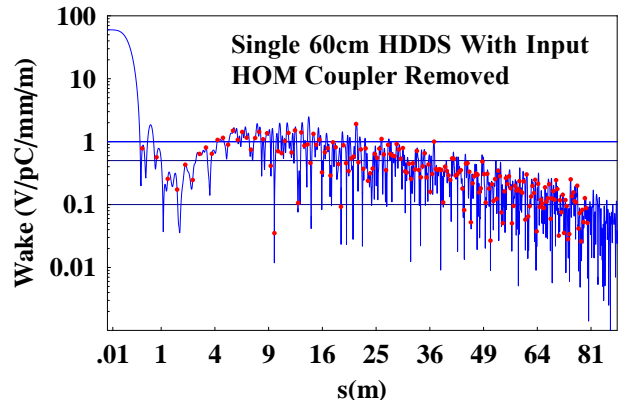
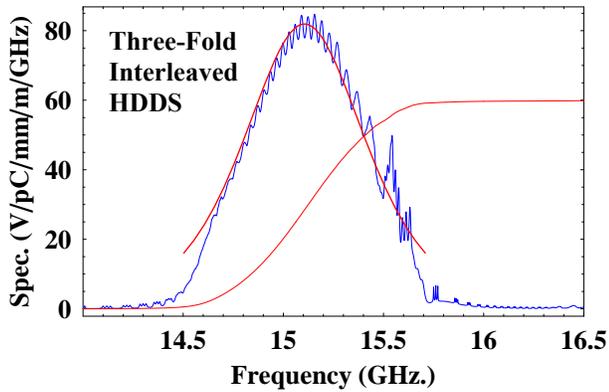
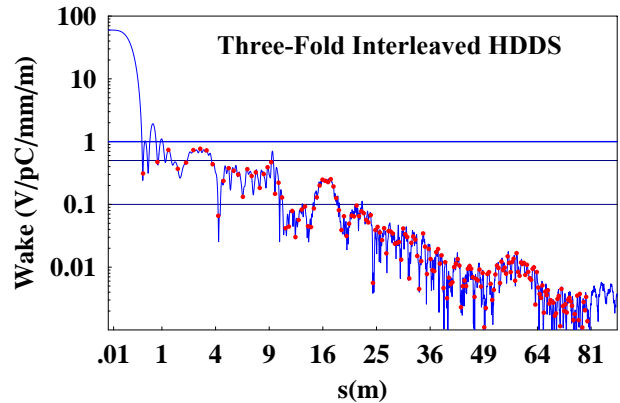

Figure 1: Spectral function for a DDS version of H60VG3 under three different conditions. Shown uppermost is the spectral function for a single 60cm structure in which the cost function has been minimized. The bandwidth of the frequency distribution is 8.40% and a total frequency width of 3.44 σ. The middle curve illustrates the effect of removing the upstream fundamental mode coupler. Becasue there are only 55 cells then large oscillations in the spectral function occur. The lowermost curve is that of the spectral function for three-fold interleaving of the frequencies of adjacent structures. The resulting optimized distribution has a bandwidth of 7.98% and a frequency width of 3.5σ.

Figure 2: Envelope of wake function for the conditions given in Fig. 1. The abscissa is chosen so that the wake is a function of the square root of the displacement behind the beam. This functional dependence magnifies the short range region. Beam dynamics simulations have shown that the wake must be below 1 V/pC/mm/m in order that BBU be prevented and this is clearly not the case for non-interleaved structures. The lower, three-fold interleaved wake is below unity at the position of every bunch (marked with red dots).

relatively unaffected by removal of the upstream higher order mode coupler but this is clearly not this case for this new short structure. In these designs we chose an unperturbed kick factor (K) weighted density function [4] of the form:

$$Kdn/df = \text{sech}^{1.5}(f/\sigma) \quad (2.2)$$

where f is the dipole frequency of each cell. This results in a slightly improved wakefield compared to the Gaussian distribution we have used in other designs. We chose to minimise the "cost" function, defined as the sum of the squares of the standard deviation and rms of the sum wakefield (defined as the sum of the wakefield of all previous bunches up until the bunch under consideration). The lowermost wakefield in Fig 2, is obtained for 3-fold interleaving, and it clearly illustrates the utility of the

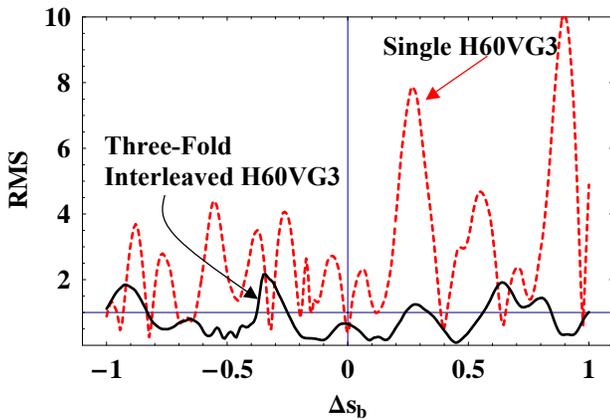

**F**igure 3: RMS of sum wakefield versus fractional change in the difference of the bunch spacing from the nominal spacing of 0.42cm minimisation code which has resulted in the first bunch being placed at the first minimum of the envelope of the wake function and all others at positions which minimise the cost function.

The rms of the sum wake provides an indication as to whether or not BBU (Beam Break Up) will occur and it is shown in Fig. 3 as the bunch spacing is varied (corresponding to a systematic error in the cell frequencies). Past experience has indicated that provided the rms is less than unity then BBU does not occur and this is the case for three-fold interleaved H60VG3 to within +/-30MHz of the central dipole frequency. The standard deviation of the sum wakefield provides information as the alignment tolerance of the structures. In the following section we calculate these tolerances in detail.

## 3. ALIGNMENT TOLERANCES

The transverse long range and short range wakefield dilutes the final emittance of the beam at the end of the linac and it imposes a tolerance on the alignment of groups of cells. The tolerance due to the long range transverse wakefield is calculated both with an analytical method [6] and by numerically tracking the beam down the linac and moving groups of cells transversely in a random manner (for a specified rms offset) with the computer code LIAR[7]. The tolerance under the influence of short range wakes is obtained entirely by tracking through the full linac. The tolerances that result from this procedure for a maximum emittance dilution of 10% are shown in Fig 4 for the linac parameters given in [6], modified to take into account 3-fold interleaving of structures (an initial Gaussian, *non-optimised* and 3-fold interleaved H60VG3 structure is used). The agreement between the analytical model and the tracking method is very good. The long-range wake imposes a single-cell alignment tolerance of 20μm and a structure-to-structure alignment tolerance of 42μm.

The short-range wake imposes a tolerance that is proportional to the square root of the length of the section under consideration and it is clear from Fig 5 that the structure-to-structure tolerance is dominated by this effect. However, we note that beam-based alignment is used to enable this tolerance to be met.

For the optimised distributions, shown in section 2, we expect the tolerances to be relaxed considerably. These tolerances will be provided in a later publication.

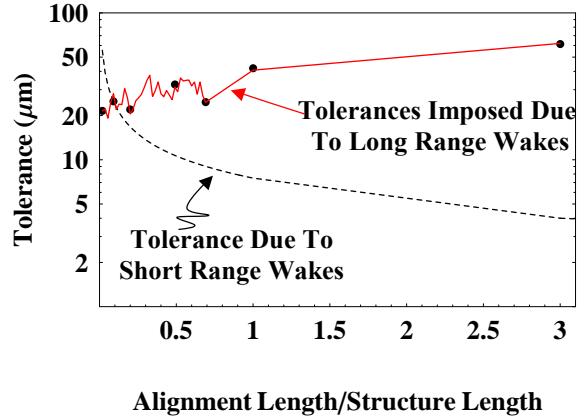

**F**igure 4: Tolerance imposed by transverse wakefields. The solid line is obtained from an analytical formula and results purely from the influence of long-range transverse wakefields. The dots are obtained by particle tracking. Also shown, with a dashed line, is the tolerance based on short-range wakes and this is also obtained by tracking the progress of the beam down the linac.